\documentstyle [12pt,a4p,epsfig,amsmath,multicol]{article}
\begin{document}
\vspace*{0.6cm}

\begin{center} 
{\normalsize\bf Spatially-separated synchronised clocks in the same inertial frame:
   Time dilatation, but no relativity
  of simultaneity or length contraction}
\end{center}
\vspace*{0.6cm}
\centerline{\footnotesize J.H.Field}
\baselineskip=13pt
\centerline{\footnotesize\it D\'{e}partement de Physique Nucl\'{e}aire et 
 Corpusculaire, Universit\'{e} de Gen\`{e}ve}
\baselineskip=12pt
\centerline{\footnotesize\it 24, quai Ernest-Ansermet CH-1211Gen\`{e}ve 4. }
\centerline{\footnotesize E-mail: john.field@cern.ch}
\baselineskip=13pt
\vspace*{0.9cm}
\abstract{ The Lorentz transformation is used to
 analyse space and time coordinates corresponding to two
  spatially-separated clocks in the same inertial frame. The time dilatation effect
 is confirmed, but not `relativity of simultaneity' or `relativistic length contraction'.
  How these latter, spurious, effects arise from misuse of the Lorentz transformation
  is also explained.}
 \par \underline{PACS 03.30.+p}
\vspace*{0.9cm}
\normalsize\baselineskip=15pt
\setcounter{footnote}{0}
\renewcommand{\thefootnote}{\alph{footnote}}

    The Lorentz transformation (LT) relates space-time coordinates ($x$,$y$,$z$,$t$) of an
  event in one inertial frame, S, to those of the same event ($x'$,$y'$,$z'$,$t'$) in another
  inertial frame S'.  As is conventional, it is assumed that S' moves with uniform velocity,
  $v$, along the common $x$,$x'$ axis of the two frames, and only events lying along the
  $x,x'$ axis are considered. 
  \par In the following, it is assumed that the time $t$ is registered by a clock
    at some fixed position in S. This records the same time, at any instant, as any
    synchronised clock at rest in S, at a different position, that may be compared locally with
   the reading of a moving clock. Two other clocks, C1 and C2, registering
   times $t'_1$ and  $t'_2$ are at rest in S' on the $x'$-axis at $x'_1 = 0$ 
   and $x'_2 = L'$ respectively. The equations of motion of these clocks in S are
    $x_1 = v t_1$ and $x_2 = v t_2+L$ respectively, where $L$ is the (time and velocity
     independent) separation in S of C1 and C2 any given instant $t = t_1 = t_2$.
      \par The space-time LT for events on the world line of C1 is
   \begin{eqnarray}
      x'_1 & = & \gamma [x_1 - v t_1] = 0 \\ 
      t'_1 & = & \gamma [t_1 - \frac{v x_1}{c^2}]
   \end{eqnarray}
     while that for C2 is:
  \begin{eqnarray}
      x'_2-L' & = & \gamma [x_2-L - v t_2] = 0 \\
      t'_2 & = & \gamma [t_2 - \frac{v(x_2-L)}{c^2}]
   \end{eqnarray}
    where $\gamma \equiv 1/\sqrt{1-(v/c)^2}$. 
     \par Eqs.~(1)-(4) are written in such a way that when $t_1 = t_2 = t = 0$ then
     $t'_1 = t'_2 = t' = 0$ so that, at this instant, C1 and C2 are synchronised.
      The constant $L = x_2(t_2 = 0)$ corresponds to a particular ($v$-independent)
     choice of spatial coordinate system in the frame S. Since as $v \rightarrow 0$, 
  ${\rm S} \rightarrow {\rm S'}$ and  $x \rightarrow x'$, Eq.~(3) becomes, when $v = 0$
   \begin{equation}
    x'_2 - L' =  x'_2 - L = 0
    \end{equation}
    from which it follows that $L' = L$. There is therefore no relativistic
      `length contraction' (LC) effect. 
     \par Using (1) to eliminate $x_1$ from (2) and (3) to eliminate  $x_2$  from
      (4) gives the time dilatation (TD) relations:
   \begin{eqnarray}
      t_1 & = & \gamma t'_1 \\
 t_2 & = & \gamma t'_2
    \end{eqnarray} 
     Therefore, at any given instant in the frame S when  $t_1 = t_2 = t$, then
     $t'_1 = t'_2$ and the clocks C1 and C2 remain synchronised ---there is no
     relativity of simultaneity' (RS) effect.
     \par The TD relations (6) and (7), when $t_1 = t_2 = t$, may be compared to
      their Galilean limit where $c \rightarrow \infty$: $t =  t'_1 = t'_2$, The TD relations
       are the only modifications of space time physics due the Lorentz
        transformation; the space transformation equations (1) and (3) become, when $\gamma = 1$,
       those of Galilean relativity, which give the same equations of motion $x'_1 = 0$, $x'_1 = vt_1$ and
        $x'_2 = L$, $x'_2 = vt_2+L$, of C1 and C2 respectively, in S',S, as Eqs.~(1) and (3).
    \par As previously discussed in detail elesewhere~\cite{JHFLLT,JHFCRCS,JHFACOORDS}, the spurious
     RS and LC effects
    of conventional special relativity are the result of an incorrect use of the space-time
       LT
    to analyse space and time measurements. For example, the coordinates of C2 are simply substituted into
       the LT (1) and (2) appropriate for the clock C1 giving:
        \begin{eqnarray}
      x'_2 = L'& = & \gamma [x_2 - v t_2]  \\ 
      t'_2 & = & \gamma [t_2 - \frac{v x_2}{c^2}]
   \end{eqnarray}
            Setting $t_2 = 0$ in (8) gives
      \begin{equation}
       L' =  \gamma x_2(t_2 = 0) = \gamma L
    \end{equation}
       while using (8) to eliminate $x_2$ from (9) gives:
     \begin{equation}
         t'_2  = \frac{t_2}{\gamma} - \frac{\gamma v L}{c^2}
    \end{equation}
      in contradiction to the translationally-invariant TD relation of 
      Eq.~(7). Eqs.~(10) and (11) are typical of text-book `derivations' of the LC and
      RS effects respectively. The former was given in Einstein's
      original special relativity paper~\cite{Ein1}.
    \par The LT (3) and (4) with $L' = L$ may be written as:
   \begin{eqnarray}
      x'_2 & = & \gamma [x_2 - v t_2]+X(L)  \\ 
      t'_2 & = & \gamma [t_2 - \frac{v x_2}{c^2}]+T(L)
   \end{eqnarray}
    where $X(L)$ and $T(L)$ are the constants:
\begin{eqnarray}
   X(L) & \equiv & (1-\gamma)L \\
  T(L) &  \equiv & \frac{\gamma v L}{c^2}
  \end{eqnarray}
   The necessity to add such constants to the right sides of (1) and (2) in order
   to correctly describe synchronised clocks at different spatial positions
   was aleady pointed out by Einstein, though, to the present writer's
   best knowledge, this was never done, either by Einstein himself or other authors,
   before the work presented in Ref~\cite{JHFLLT}. The important passage occurs
     in \S 3 of Ref~\cite{Ein1}, immediately after the derivation of the LT
     as in (1) and (2) above. In the English translation of Perrett and Jeffrey it is:
   `If no assumption is made as to the initial position of the moving system and as to
    the zero point of $\tau$' ($\tau$ is $t'$ in the notation of the present letter) `an additional
    constant is to be placed on the right side of each of these equations.' (`these equations'
      are equivalent to Eqs.~(1) and (2) of the
   present letter).

    \par The essential point made here is that physics, either that underlying
     the clock mechanism, or the relativistic TD effect of Eqs.~(6) and (7),
     predicts only the observed {\it rate}
     of a moving clock, not its setting. The spurious RS and LC effects arise because
    a clock {\it setting}, built into the `standard' LT (1) and (2),
    is misinterpreted as a physical {\it time difference} between the readings of
    sychronised clocks at a different spatial
    positions in an inertial frame, when they are viewed by an observer 
    in a different inertial frame.  .  
    \par  Experiments have recently been proposed to
    search for the existence (or not) of the RS effect~\cite{JHFSEXPS}. At the time of writing,
    there is ample experimental verification of TD but none of RS or LC~\cite{JHFLLT}. 

\end{document}